\documentclass{elsart}
\usepackage{graphics}
\begin{document}
\begin{frontmatter}
\title{Lessons from Hadron Phenomenology  }
\author{ M.M. Brisudova$^{\dagger}$, L. Burakovsky$^{\ddagger}$ and 
T. Goldman$^{\ddagger}$ }
\address{$^{\dagger}$ Physics Dept., University of Florida, Gainesville, FL 
32611 \\ \hspace*{0.1cm} $^{\ddagger}$ Theoretical Division, T-16, LANL, 
Los Alamos, NM 87544 }
\date{  }
\begin{abstract}
Meson spectra can be well approximated by a specific form of a nonlinear Regge 
trajectory which is consistent with a finite number of bound states. This may 
have important consequencies for experiment, and may be a hint for the theory.
\end{abstract}
\end{frontmatter}

\section{Introduction}
 
The main effort of our light-front community has been in the past ten years to 
develop new techniques for solving QCD, be it DLCQ- or renormalisation-group- 
based methods (for review see \cite{REVIEW}). 
In practice the usefulness of any procedure (which includes the question of 
convergence) is often determined by how well its first approximation captures 
the essential features of the system under consideration.
This is why phenomenology can be helpful in tailoring the approximation 
schemes by giving us some hints of what is important for the spectra.
 
\section{Regge Trajectories and Spectroscopy}
One phenomenological way of describing the spectra is via Regge trajectories 
$\alpha(s) $, which can be, for the purpose of my presentation, thought of as 
``lines'' connecting the bound states 
plotted in the angular momentum $J$ vs. mass squared $M^2$ plane, 
$\alpha(s) \equiv J(M^2).$
 Table I illustrates the grouping into trajectories  of the light $I=1$ mesons.
 The trajectories differing by 
parity (such as the vector and tensor, the first line of the Table I) would be 
degenerate in a non-relativistic theory; in the full theory, they are 
near-degenerate in the bound-state region.

\begin{center}
{\begin{tabular}{c||c|c|c|c|c|c} 
{\bf $q\bar{q}$ spin} &  & {\bf S }& {\bf P}&{\bf  D }&{\bf  F}&  {\bf wave} 
 \\ \hline
 {\bf S=1} & {\it vector} & $\rho \ (1^{--})$ &  & $\rho_3 \ (3^{--})$ & & \\
 &  {\it tensor}&  & $a_2 \  (2^{++})$ &   &  $a_4 \  (4^{++})$ & \\ \hline
& {\it axial vector} & & $a_1 \ (1^{++})$ &  & $a_3 \ (3^{++})$ &  \\
 &{\it axial tensor} & & & $\rho_2 \  (2^{--})$ &   & \\ \hline
& {\it scalar}& & $a_0 \ (0^{++})$ &  & $a_2 \ (2^{++})$ &  \\
 & {\it vector} & & & $\rho \  (1^{--})$ &   & \\ \hline
{\bf S=0} & {\it pseudoscalar} & $\pi \ (0^{-+})$ &  & $\pi_2 \ (2^{-+})$ & &
 \\
 & {\it pseudovector} &  & $b_1 \  (1^{+-})$ &   &  $b_3 \  (3^{+-})$ & \\ 
\hline
\end{tabular}
}
\vskip0.2in
\end{center}
{\bf Table I.} Regge trajectories for the light $I=1$ mesons. $J^{PC}$ of the 
given meson is shown in parentheses. \\

 In the Veneziano model for scattering amplitudes \cite{Ven} there are 
infinitely many excitations populating linear Regge trajectories. 
The Veneziano amplitude ($s,$ $t$ are the usual Mandelstam variables), 
\begin{equation}
V(s,t) 
= \frac{\Gamma(1-\alpha(s))\Gamma(1-\alpha(t))}{\Gamma(2-\alpha(s)-\alpha(t))}
=\int _0^1 dx \, x^{-\alpha(s)} (1-x)^{-\alpha (t)},
\label{veneziano}
\end{equation}
only allows for a linear (or at most a polynomial \cite{Thorn}) trajectory, 
$\alpha(s) = \alpha (0) + \alpha ' s$, if the spins are finite \cite{Ven}.
The same picture of linear trajectories arises from a linear confining 
potential 
and the string model of hadrons (see \cite{us} for references).

However, the realistic Regge trajectories extracted from data are {\it 
nonlinear.} For example, 
the straight line which crosses the $K_2^\ast $ and $K_4^\ast $ squared masses 
corresponds to an intercept $\alpha _{K_2^\ast }(0)\approx 0.1,$ whereas the 
physical intercept is located at $\approx 0.4.$ In addition to being 
disfavored by this and other experiments, linear trajectories also 
lead to problems in theory. (For details on both experimental and theoretical 
aspects, and references, see \cite{us}.)

We have argued that the nonlinearity of hadronic Regge trajectories is due to 
pair production which screens the confining QCD potential at large distances 
\cite{us0}. Because of pair production, there cannot be infinitely many bound 
states, and the real part of Regge trajectories will terminate in angular 
momentum. Once the nonlinearity of Regge trajectories is an accepted fact, 
the question of what specific form should be used for phenomenology arises.

We have considered a whole class of nonlinear trajectories allowed by dual 
amplitudes with Mandelstam analyticity (DAMA) (\cite{Jenk} and references 
therein). DAMA are a generalization of Veneziano amplitudes 
(\ref{veneziano}),
\begin{eqnarray}
D(s,t) & = & \int _0^1 dx \, \left({x\over{g}}\right)^{-\alpha(s')} 
\left({1-x\over{g}}\right)^{-\alpha (t')}
\label{dama}
\end{eqnarray}
where $g>1$ is a constant, and $s'=s(1-x),$  $t'=t x$, and they allow for 
Regge trajectories of a form
\begin{eqnarray}
\alpha_{j\bar{i}} (t) & = & \alpha _{j\bar{i}}(0)+\gamma \Big[ 
T_{j\bar{i}}^\nu -(T_{j\bar{i}}-t)^\nu \Big]   
\label{trajectory}
\end{eqnarray} 
 where $\nu$ is a constant restricted to $0 \leq \nu \leq {1\over{2}}$; 
$\gamma $ is a universal constant; $T_{j\bar{i}}$ is a trajectory threshold, 
$\alpha_{j\bar{i}}(0)$ is its intercept, and $i, j$ refer to flavor. 
We have argued that both limiting cases ($\nu =1/2$ and $\nu=0$) can be 
expected to work comparably well for lowest lying states, but the $\nu=1/2$, 
so-called square-root, form  is likely to be more realistic. 
Therefore, we use the square-root form for phenomenological purposes.

Assuming that mesonic Regge trajectories are of the  form (\ref{trajectory}) 
with $\nu =1/2$, we determine thresholds and intercepts of trajectories by using 
various 
experimental information. Typically, we use masses of a few known lowest lying
states, and in the case of the $\rho$ trajectory we also use the value of the 
intercept (which is known and well-established) found from exchange processes.
The value of $\gamma$ (the universal asymptotic slope) is fit to the $\rho$ 
trajectory, and then taken as universal for all other trajectories.

The approach has more predictive power than one would naively expect. This is 
because the parameters for trajectories with different flavors within a 
multiplet are related by 
additivity of intercepts 
and
additivity of inverse slopes near the origin,
two requirements that are independent of which specific form is assumed for 
the trajectories \cite{us}. This means in practice, that out of 20 parameters 
for each 5-flavor meson multiplet, only 8 are independent. We further reduce 
the number of parameters by requiring that the thresholds of parity-partner 
trajectories (such as vector and tensor in Table I) are the same. This further 
reduces the number of parameters for the two related multiplets from 16 to 12. 
We tested this assumption where there were data available, and it is well 
satisfied \cite{us}.

Once the parameters of the trajectories are known,   
masses of excited states lying on each of the trajectories under consideration 
can be calculated, and compared with experimental data. 
 Where there are no data available, our results are predictions. Recall, 
that one of the parameters of a trajectory, in particular, its intercept, 
is also an observable, so the result of our fit to bound state data can 
be independently checked with what is known from scattering processes. 

We studied vector and  tensor meson trajectories, and pseudoscalar and 
axial-vector meson trajectories. In all of these cases, our results are in 
excellent agreement with both scattering and bound state data, as 
well as results of different methods, such as lattice QCD or sum rules.

The approach can also be applied to glueballs. We extracted parameters of the 
Pomeron trajectory, assuming that it, too, is of the square-root form, from 
the data collected by ZEUS collaboration \cite{ZEUS}. We found a threshold 
of 10 GeV, much higher that that of, but consistent with what can be expected 
from, light mesons. The parameters yield plausible values for glueball masses 
\cite{usf}.

Experimental consequences of the termination of Regge trajectories are 
intriguing. For example, we believe that $a_6$ is likely to be the last state 
on the $a_2$ trajectory. We also believe that any state above roughly 3.2 GeV 
that does not fit into charmonium spectra is likely to be an exotic, since 
according to our study light quarkonium states terminate at about 3 GeV. This 
leaves a large window for glueballs, since they terminate at a much higher 
mass. Of no less interest is the main lesson for the theory: Linear 
confinement is likely neither sufficient nor the most important factor 
determining 
the position of poles, not even for states as low as second or third on a 
trajectory. This means that if our light-cone attempts to solve QCD are to be 
successful beyond the ground states, they need to take into account the effect 
of pair production or, in other words, mixing with higher Fock states.
 
M.B. would like to thank  
Antonio Bassetto for discussions. 
This research is supported in part by the U.S.D.O.E.   
under contract W-7405-ENG-36, and grant DE-FG02-97ER-41029. 
  
\end{document}